\pdfoutput=1
\documentclass[aps,pra,floatfix,twocolumn,reprint]{revtex4-1}
\usepackage{graphicx}
\usepackage{bm}
\usepackage{mathrsfs}
\usepackage{amssymb}
\usepackage{amsthm}
\usepackage{amsfonts}
\usepackage{amsmath}

\begin{document}
\title{Learning a compass spin model with neural network quantum states} 
\author{Eric Zou}
\affiliation{Department of Physics and Astronomy, George Mason University, Fairfax, Virginia 22030, USA}
\author{Erik Long}
\affiliation{Department of Physics and Astronomy, George Mason University, Fairfax, Virginia 22030, USA}
\author{Erhai Zhao}
\affiliation{Department of Physics and Astronomy, George Mason University, Fairfax, Virginia 22030, USA}

\begin{abstract}
Neural network quantum states provide a novel representation of the many-body states
of interacting quantum systems and open up a promising route to solve frustrated quantum spin
models that evade other numerical approaches. Yet its capacity to describe complex
magnetic orders with large unit cells has not been demonstrated, and its performance in
a rugged energy landscape has been questioned. Here we apply restricted Boltzmann machines and stochastic 
gradient descent to seek the ground states of a compass spin model on the honeycomb lattice, which unifies the Kitaev model, Ising model
and the quantum 120$^\circ$ model with a single tuning parameter. 
We report calculation results on the variational energy, order parameters and correlation functions.
The phase diagram obtained is in good agreement with the predictions of tensor network ansatz, demonstrating the capacity of 
restricted Boltzmann machines in learning the ground states of frustrated quantum spin Hamiltonians.
The limitations of the calculation are discussed. A few strategies are outlined to address some of the challenges 
in machine learning frustrated quantum magnets.
\end{abstract}
\maketitle

\section{Introduction}
Finding the ground-state wave functions of frustrated quantum spin models \cite{fru1,fru2} in two dimensions (2D)
remains an outstanding theoretical challenge despite the great strides made in recent decades 
in the field of numerical many-body algorithms \cite{spin-liquid-review}. 
On the one hand, exact diagonalization 
and density matrix renormalization group (DMRG), when applied to 2D, are restricted by the finite system size. 
On the other hand, variational Monte Carlo approaches, while being unbiased and valid in the 
thermodynamic limit, 
depend on the quality of the trial wave functions and tend to lose its prediction power due to
the lack of convergence associated with the prevalent ``negative sign problem" in frustrated quantum spin systems.
Another efficient variational ansatz for quantum spin models is based 
on the tensor network (TN) representation of the many-body wave functions \cite{TN1,orus2014}. 
One type of tensor networks known as projected entangled pair states, which generalize
the matrix product states from one dimension to two dimensions, have been successfully applied to quantum 
spin models \cite{tensor-spin}. The accuracy of TN ansatz however depends on the approximations
employed when truncating and contracting the tensors. 
Given that each method has its advantages as well as drawbacks, a holistic approach will benefit from new
numerical methods that can shed fresh light on this persistent problem.

Inspired by its tremendous success in machine learning and artificial intelligence \cite{mehta-rev}, 
neural networks were recently proposed to solve quantum spin models \cite{carleo-troyer,rbm-rev}. 
A generic wave function of $N$ interacting spins, say $S=1/2$, is a superposition of $2^N$
basis states with complex coefficients. Mathematically, the wave function $\psi(\mathbf{s})$
defines a mapping from vector $\mathbf{s}=(s_1, s_2, ..., s_N)$,
where $s_i=\uparrow$ or $\downarrow$, to a complex number. Therefore, it can
be thought of as a machine that gobbles up  $\mathbf{s}$ and spits
out a complex number. 
It is conjectured that such mapping can be represented 
accurately by {neural networks with sufficient number} of nodes, layers, and connections \cite{bengio-repres}. 
Then, to find the ground state of a given interacting spin Hamiltonian, all one needs to do is to train 
the network by adjusting its parameters stochastically so that 
the energy expectation value is minimized.
This approach was pioneered by
Carleo and Troyer, who represented the many-spin wave function as restricted Boltzmann machines (RBMs) 
and successfully applied it to solve the Heisenberg model on the square lattice \cite{carleo-troyer}.
In addition to RBM, the wave function can also be expressed as feed-forward neural networks 
or other neural network architectures. 
For example, Choo et al employed convolutional feed-forward neural networks to solve the $J_1$-$J_2$ model, 
a canonical example of frustrated magnets believed to host a quantum spin liquid,
and obtained excellent energetics comparable to  
exact diagonalization and DMRG \cite{choo-carleo-J1J2}. The $J_1$-$J_2$ model has also been 
investigated in Refs. \cite{He-J1J2,cai,sign2,nomuraprx,brian} using the neural network ansatz.
In a broader context, recent work has revealed a remarkable connection between 
the neural-network and the tensor-network representation of quantum many-body states 
despite their differences in appearance and origin \cite{cirac-NN-TN,xiang-equiv,moore-TN}.
It was shown that neural network quantum states can describe states with
topological order, even with entanglement entropy beyond the area law 
\cite{deng-entangle,deng-topo,chiral}. Beyond quantum spin models, 
neural network quantum states have been also applied to strongly correlated 
fermions \cite{imada,melko-fermion,khatami}.

These promising developments have raised many open questions.
So far the neural network ansatz was able to identify relatively simple states such as the Neel
or stripe order. Is it capable of finding more exotic phases with complicated symmetry breaking patterns? 
How can it be applied to extract the entanglement signatures of quantum spin liquids? 
Does it provide an accurate, practical method to determine the phase transitions  
by computing the order parameters and correlation functions? 
Very recently, certain limitations to the expressive power of RBMs as well as 
the stochastic reconfiguration algorithm have been noted. For example, in some cases, the algorithm suffers 
from inherent numerical instabilities \cite{sign1,sign2}.
For highly frustrated quantum spin models, e.g. near $J_2/J_1=0.5$ in the $J_1$-$J_2$ model,
the energy landscape is believed to be rugged, the approach to the global minimum may not be guaranteed in
practice. These limitations led to ongoing efforts to represent the amplitude and the phase of the wave function 
separately using two real-valued networks, and to learn the sign structures of the wave function to 
facilitate the convergence \cite{sign2}. 

In light of these open questions, in this work we apply the neural network ansatz to the tripod model \cite{npj-tripod},
a frustrated
quantum spin model in two dimensions. It contains the Kitaev model \cite{kitaev} as a special limit and has an extended 
spin liquid phase. At the same time, its phase diagram also includes the Neel order and a nontrivial
bond order which, according to TN calculations \cite{npj-tripod}, can be viewed as a periodic lattice of spin vortices. Thus, this model provides an ideal 
playground to test the performance and limitations of the neural network ansatz.
We note that previously, there have been several works that applied neural networks to study the ground state
and excitations of the Kitaev model or its generalizations, e.g. with external magnetic field or Heisenberg terms \cite{kita-rbm,rbm-MC,xu-21,lode}. 
The model here is rather different: it overlaps with the Kitaev model only at one special point. Moreover, our
primary focus is on the phase diagram and phase transitions between the spin liquid and the long-range ordered
states.
 
This paper is organized as follows. In section II, we introduce the tripod model and summarize existing numerical results
from tensor network ansatz. Then we outline the RBM ansatz in section III. Section IV
gives a detailed discussion of our main numerical results, including the energy, the order parameters, and the resulting phase
diagram. In section V, we discuss the limitations of the neural network ansatz as implemented in our work, and 
directions for future improvement. We hope our results, including the strategies employed to facilitate 
the learning process, can be useful for applying the neural network ansatz to other quantum spin models, and more generally, to quantum
many-body systems.

\section{The tripod model}
The tripod model is a quantum spin model defined on the two dimensional honeycomb lattice.
It belongs to compass spin models, a broad class of Hamiltonians in which the exchange interaction
between two neighboring spins
depends on the spatial direction of the bond. The study of compass models 
has a long history, for review see Ref. \cite{compass}. Perhaps the best 
know example is the Kitaev model \cite{kitaev}: along the three bond directions of the honeycomb lattice,
the spin exchange interaction is given by $S_xS_x$, $S_yS_y$, and $S_zS_z$ respectively 
(in this shorthand notation, 
the first spin operator is for one lattice site
and the second for a neighboring site). 
Another interesting example of compass models is the quantum 120$^\circ$ model discovered by Zhao and
Liu \cite{ez-120}, and independently by Wu \cite{wu-120}, in the study of strongly interacting $p$-orbital fermions. 
In this model, the 
spin exchanges along the three bonds of the honeycomb lattice are given by 
$S_1S_1$, $S_2S_2$, and $S_3S_3$ respectively. In spin space,
each spin operator is represented by a vector, and here the three spin vectors $S_{1,2,3}$ lie
within a plane forming 120$^\circ$ angle with each other. It is apparent that the 120$^\circ$ model 
is a cousin of the Kitaev model where the three corresponding spin vectors $S_{x,y,z}$ form an orthogonal triad
in spin space (i.e.  90$^\circ$ angles with each other).
This intimate connection between the two models motivated the authors of Ref. \cite{npj-tripod} to 
unify the 120$^\circ$ model, the Kitaev model, and the Ising
model into a single compass model
parameterized by an angle $\theta$. Its Hamiltonian is given by
\begin{equation}
H =J\sum_{\mathbf{r},\gamma}S_\gamma(\mathbf{r}) S_\gamma({\mathbf{r}+\mathbf{e}_\gamma}).
\label{eq:Ham}
\end{equation}
Here $J>0$ is the antiferromagnetic coupling, $\mathbf{r}$ labels the lattice site,
and $\mathbf{e}_{\gamma}$ with $\gamma=1,2,3$ denotes the three bond vectors of the honeycomb lattice,
i.e. $\mathbf{r}+\mathbf{e}_\gamma$ is a neighboring site of $\mathbf{r}$ (we have set the 
lattice spacing to one).
The spin 1/2 operator $S_\gamma$ is defined as
\begin{equation}
S_\gamma =\frac{1}{2}(\tau_z\cos\phi_\gamma+\tau_x\sin\phi_\gamma)\cos\theta+\frac{1}{2}\tau_y\sin\theta
\end{equation}
where  $\tau_x, \tau_y, \tau_z$ are the Pauli matrices, and $\phi_\gamma=0,2\pi/3,4\pi/3$
are the azimuthal angle of the corresponding bond direction $\mathbf{e}_\gamma$.
For brevity, we have suppressed the $\theta$ dependence
of $S_\gamma $ and $H$, and the notation $S_\gamma(\mathbf{r})$ means the spin operator $S_\gamma$ 
is localized at site $\mathbf{r}$.  

Model Eq. \eqref{eq:Ham} is dubbed the tripod model, because 
geometrically the three $S_\gamma$ form a tripod in the spin space as shown in Fig. 1 of Ref. \cite{npj-tripod}.  They are tilted out of the $xz$ plane by angle $\theta$ and, when projected onto the $xz$ plane, are 120$^\circ$  from each other.
In addition to the tilting angle $\theta$ that defines $S_\gamma$, it is convenient to follow Ref. \cite{npj-tripod} to introduce $\theta'$, the angle
between $S_1$ and $S_2$, i.e. the angle subtended by the two adjacent legs of the tripod. 
The two angles are related by 
$\cos\theta'=1-({3}/{2})\cos^2\theta$. Three limits can be identified as we change $\theta$ from 0 to  90$^\circ$. 
At $\theta=0$ (and correspondingly $\theta'=120^\circ$), the three legs of the tripod
$S_{1,2, 3}$ are fully open and lie within the
$xz$ plane. In this limit, $H$ reduces to the quantum $120^{\circ}$ model.
As $\theta$ is increased, the three legs are increasingly tilted out of the $xz$ plane, corresponding to
a  partially open tripod.
At {$\theta=\theta_K=\mathrm{arccos}(\sqrt{2/3})\simeq 35.26^\circ$}, $\theta'$ becomes exactly $90^\circ$, then the tripod model becomes the Kitaev model:
now that the three operators $S_\gamma$ are orthogonal to each other, we can carry out a spin rotation
and redefine them as $S_{x,y,z}$. Finally, when $\theta$ is increased all the way to 90$^\circ$,
we have $\theta'=0$ and all three $S_\gamma$ collapse to the $y$ axis. The tripod is now fully closed.
In this limit, $H$ reduces to the Ising model,
$H=({J}/{4})\sum_{\mathbf{r},\gamma} \tau_y(\mathbf{r}) \tau_y({\mathbf{r}+\mathbf{e}_\gamma})$.
Note that usually the Ising interaction is written in the form of $\tau_z\tau_z$. Here to make it easier to compare with
previous literature, we follow the convention to choose $\tau_y$ as the vertical axis in spin space \cite{npj-tripod}.
This choice of the axes is particularly convenient to recover the 120$^\circ$ model 
defined in earlier work Ref.~\cite{ez-120}. In passing, we note that the tripod model is not only of
theoretical interest due to its synthesis of three important models in quantum magnetism. 
Recent experiments on honeycomb antiferromagnet NaNi$_2$BiO$_{6-\delta}$ 
suggest that its dominant exchange interactions resemble those in the tripod model
with additional terms, such as the Heisenberg exchange, also playing a role \cite{neutron}.

Some limits of the tripod model are easy to understand. For example, 
in the Ising limit, the ground state has Neel order, and the order parameter 
is the staggered magnetization along $y$. 
At the Kitaev point, the model is analytically solvable, and its ground state is a spin liquid and has no long-range
magnetic order \cite{kitaev}. Aside from these two limits,
for general $\theta$, the tripod model must be solved numerically.  This is challenging because the model
is frustrated and hosts highly nontrivial quantum phases. In particular, 
the ground state of the $120^{\circ}$ model has been somewhat controversial \cite{ez-120,wu-120,nasu}. It was 
conjectured to be long-range ordered despite the geometric frustration \cite{wu-120}.
Ref. \cite{npj-tripod} for the first time solved the tripod model for general $\theta'$
and obtained its ground state phase diagram using tensor network ansatz.
The main conclusion is that there are three phases separated by two phase transitions,
see Fig. 1 of Ref. \cite{npj-tripod}. 
In particular it predicted that the ground state of the $120^{\circ}$ model has valence bond order.
In this state, all the spin are confined
within the $xz$ plane to form a periodic pattern which can be viewed as a triangular lattice 
of hexagons. Around each hexagon, 
the spin winds successively at a $60^\circ$ interval, forming
a discrete spin vortex, see Fig. 5 of Ref. \cite{npj-tripod}. Note that this phase was referred to as ``dimer phase" 
in Ref. \cite{npj-tripod}, because along the bonds connecting these hexagons, 
two neighboring spins point in opposite directions. Such a terminology is unconventional, because in the literature 
``dimer" is usually synonymous to spin singlet. To avoid potential confusion, we prefer to call this phase having
{\it valence bond order}, because it features spatially periodic modulations of the bond energy.
TN ansatz also predicted that the quantum spin liquid is stabilized within the finite window $\theta'\in [87^{\circ},94^{\circ}]$ enclosing the Kitaev point $\theta'= 90^\circ$. Judging from the variation of the order parameters with $\theta'$, the valence bond to spin liquid transition seems continuous, while the spin liquid to Neel transition seems to be first order \cite{npj-tripod}.

The main goal of the present work is to investigate the ground state phase diagram of the tripod model using 
an independent method. This serves two purposes. 
On the one hand, the variational calculation with neural network ansatz here provides a check for 
the TN results, especially regarding the ground state in the $120^{\circ}$ limit as well as the location and 
nature of the phase transitions. 
On the other hand, the calculation tests the capacity of the neural network ansatz by applying it to solve a frustrated quantum spin model
which has not only spin liquid but also nontrivial long-range order with an intricate spatial pattern. 
A priori, it is unclear whether these ground states, the order parameters, or phase transitions 
can be captured by the neural network ansatz. Overall, our calculation 
benchmarks the efficiency, stability, and accuracy of the neural network algorithm
by comparing to the state-of-the-art TN results. 

\section{Restricted Boltzmann Machines}
We will represent the many-spin wave function using one of the simplest neural networks,
the Restricted Boltzmann Machines (RBMs). The implementation follows the original work
of Ref. \cite{carleo-troyer}. To avoid repetition, here we only outline the main ideas. More details can be found in Ref. 
\cite{carleo-troyer} and \cite{rbm-rev}.
A restricted  Boltzmann machine is a shallow neural network with two layers, the visible layer consisting of 
$N$ nodes characterized by spin variables $s_i$ ($i=1,2,...,N$) and a hidden layer of $M$ nodes described by variables $h_j$
 ($j=1,2,...,M$).
The coupling between node $s_i$ and node $h_j$ is described by a connection weight $w_{ij}$. 
The quantum mechanical wave function takes the form of joint Boltzmann weight \cite{carleo-troyer},
\begin{align}
|\Psi \rangle &=\sum_{\{s_i\}}\psi(\mathbf{s})|\{s_i\}\rangle \nonumber \\
&= \sum_{\{s_i\}}\sum_{\{h_j\}}e^{ a_i s_i +  b_j h_j +w_{ij}s_ih_j } |\{s_i\}\rangle.
\label{wf}
\end{align}
Here, repeated indices in the exponent are summed over, $\{s_i\}=\{s_1,s_2,...s_N\}$ are all possible spin configurations
(similarly for $\{h_j\}$), the biases $a_i$ and $b_j$ as well as the connection weights $w_{ij}$ are all complex variational parameters.
Even though there is no direct intra-layer connection in a ``restricted" Boltzmann machine,
the hidden nodes induce correlations among the visible nodes. 
For real biases and connections, it is known that RBMs can represent any classical distribution to desired accuracy with sufficient numbers of hidden units \cite{bengio-repres}. 
The expressive power of complex RBMs is less known. It has been argued that a fully connected RBM
can capture entanglement bounded by volume law and hence efficiently describe the ground states of many Hamiltonians \cite{xiang-equiv}.
In our calculations, we consider a finite honeycomb lattice with $L\times L$ unit cells with periodic boundary 
conditions. Then the number of sites $N=2L^2$ and the number of bonds $N_b=3L^2$. 
The layer density ratio $\alpha=M/N$ is a tuning parameter, we find $\alpha=2$ gives satisfactory 
performance for $L=4$.

Starting from some initial guess, e.g. random values, the variational parameters are adjusted iteratively to minimize the
variation energy $E=\langle \Psi | H |\Psi\rangle$, the expectation value of the Hamiltonian Eq. \eqref{eq:Ham} for the current wave function Eq. \eqref{wf},
computed approximately by Monte Carlo sampling \cite{carleo-troyer}. 
This is done by making stochastic moves in a large parameter space based on estimating the energy gradient. This stochastic optimization
procedure is often called {\it learning}, or training the RBM. Here many mature algorithms from the machine learning
literature can be applied \cite{mehta-rev,mit}. For example, we have tested and compared several algorithms including 
stochastic gradient descent, adagrad, and adamax \cite{mit,geron-book,netket}. 
The actual computation is carried out using the powerful Netket library \cite{netket}, aided by 
custom-made routines to manipulate the variational wave functions directly.

We emphasize that while the model and learning algorithm are relatively straightforward
to set up, the actual training of the RBM with a vast parameter space is by no means a trivial task.
This is analogous to many other complex machine learning tasks: 
efficient training a neural network hinges on understanding the particularities of the model, the parameter space, and the quantity and quality of
the data etc. \cite{mit,geron-book}. For example, starting from a random configuration of the RBM, the algorithm
may lead to a quick convergence to a local minimum and stall there. 
This becomes especially problematic in regions where a few orders compete:
for example, a blind stochastic search often yields wildly fluctuating results for two neighboring parameters 
that belong to the same phase. In this case, one may find the best energetics 
by trying different optimization algorithms or starting from different initial guesses.
Even when the true ground state is being approached, the accuracy of the converged 
energy depends critically on the proper choice of the parameters such as learning rate 
and sampling batch size. 
A more serious problem is the sporadic occurrence of numerical instability, presumably
due to the parameters being complex, which may manifest as a fast runaway of the energy  
toward divergence. 
These numerical complexities complicate the task of finding the ground state phase diagram.
 {(The performance of RBM is discussed in Ref. \cite{azizi} for classical spin models)}. 
Some of the strategies we employ to alleviate these problems are discussed below in section IV.

\section{Ground-state phase diagram}

The procedure to learn the phase diagram of the tripod model is as follows.
For a given value of the tilting angle $\theta$, the restricted Boltzmann machine is started 
from random parameter values, then stochastic moves are made to lower the variational energy until
convergence is achieved. A crucial parameter here is the learning rate $r$, or step size of the stochastic
moves \cite{mit,geron-book}. For stochastic gradient descent, choosing a $r$ that is too large can easily end up with numerical
instability, while having $r$ too small may slow the learning to a crawl and trap it inside a local minimum.
The optimal value of $r$ depends on the model and the optimizer (many popular optimizers use
adaptive learning rates determined from gradient and/or momentum). Its order of magnitude is determined by trial and error,
and its value is adjusted on the fly, for example, when entering a flat energy landscape. 
When the algorithm fails to reach the anticipated energy, different optimizers or learning parameters are tried to shake things up.
If no further progress can be made, the machine is restarted.
Some states, for example the Neel state, are rather easy to reach with fast and robust convergence, {e.g.
after hundreds of iterations.}
Other states, such as the spin liquid or bond order, require many more steps for the energy to relax, { e.g. at least thousands
of iterations even with reasonable learning parameters and initial guess.} 
This is expected because of the frustration and the presence of many competing orders.
To facilitate the search for ground states in highly frustrated regimes, it is useful to start from wave functions that were learned
previously for parameters nearby and have competitive energies. 
In this case, random fluctuations are introduced to the wave function before the run, and the result must be compared
to those obtained from random, blind guesses. During the run, the statistical errors (variance or standard deviation) of the observables
are also monitored.

\begin{figure}[h]
\includegraphics[width=0.5\textwidth]{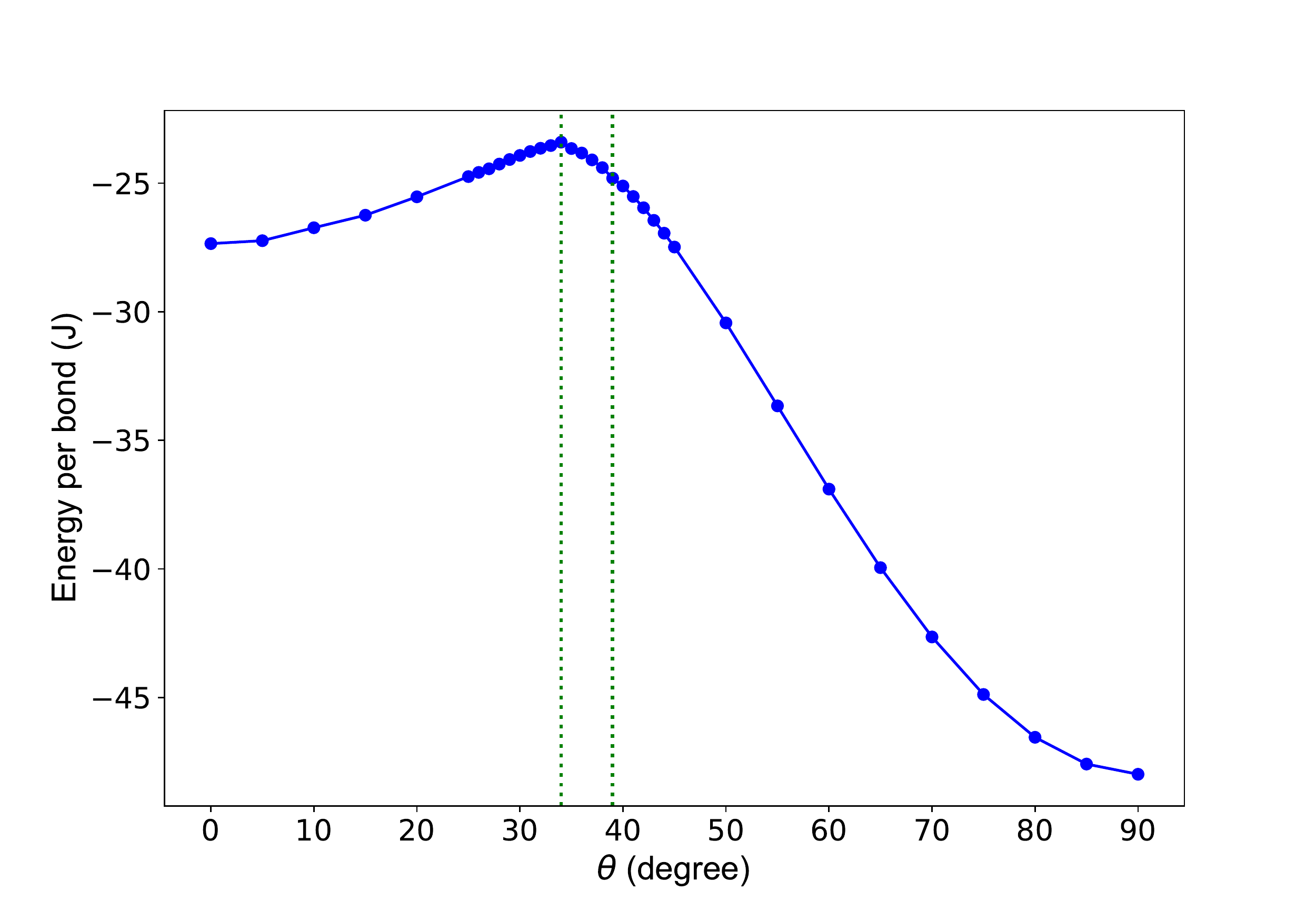}
\caption{The converged variational ground state energy per bond in units of $J$ for the tripod model Eq. (1)
obtained from Restricted Boltzmann Machines. Here $\theta$ is the tilting angle out of the $xz$ plane. 
$\theta=90^\circ$ is the Ising limit, {$\theta\simeq 35.26^\circ$} is the Kitaev point, and $\theta=0^\circ$ realizes
 the quantum 120$^\circ$ model. {The vertical dotted lines at $\theta=34^\circ$ and $39^\circ$
are guide to the eye.} The system has {32} sites with periodic boundary conditions, 
$L=4$, $\alpha=2$. No symmetry is enforced. The statistical error is smaller than the symbol size.
}\label{fig1}
\end{figure}

Fig. \ref{fig1} shows the energy per bond in units of $J$,  $\epsilon=E/N_bJ$, as a function of the tilting angle $\theta$. 
The energy is the lowest in the Ising limit $\theta=90^\circ$, and the neural network ansatz accurately reproduces
the analytical result $\epsilon=-1/4$, corresponding to the antiparallel alignment of neighboring 
spins. The convergence to ground state in this limit is rather fast, perhaps due to the classical nature of the Ising model. 
(In comparison, for $\theta<50^\circ$, reaching the ground state is not as straightforward and requires 
some of the strategies outlined above.)
As $\theta$ is reduced, $\epsilon$ rises quickly; and after going through the Kitaev point  {$\theta_K\simeq 35.26^\circ$}, it reaches its peak value 
of $\epsilon\simeq -0.122$ at $\theta=34^\circ$. The elevation in energy is in accordance with the 
fact that in this region around $\theta_K$ the system is most frustrated. 
Upon further reduction of $\theta$, the energy starts to decrease.
The noticeable cusp in energy located at $\theta=34^\circ$ marks the transition to the bond ordered phase,  
inside which the frustration is relieved to some degree but not entirely. 
This trend continues until the 120$^\circ$ model limit is reached at $\theta=0$. 
Note the spin liquid to Neel transition is not obvious by inspecting the energy alone.
But plotting $d\epsilon/d\theta$ reveals a sudden change at $\theta\sim 40^\circ$.
In what follows, we present a better way to reveal the phase boundaries by computing the order parameters 
and spin correlation functions.

\begin{figure}[h]
\includegraphics[width=0.5\textwidth]{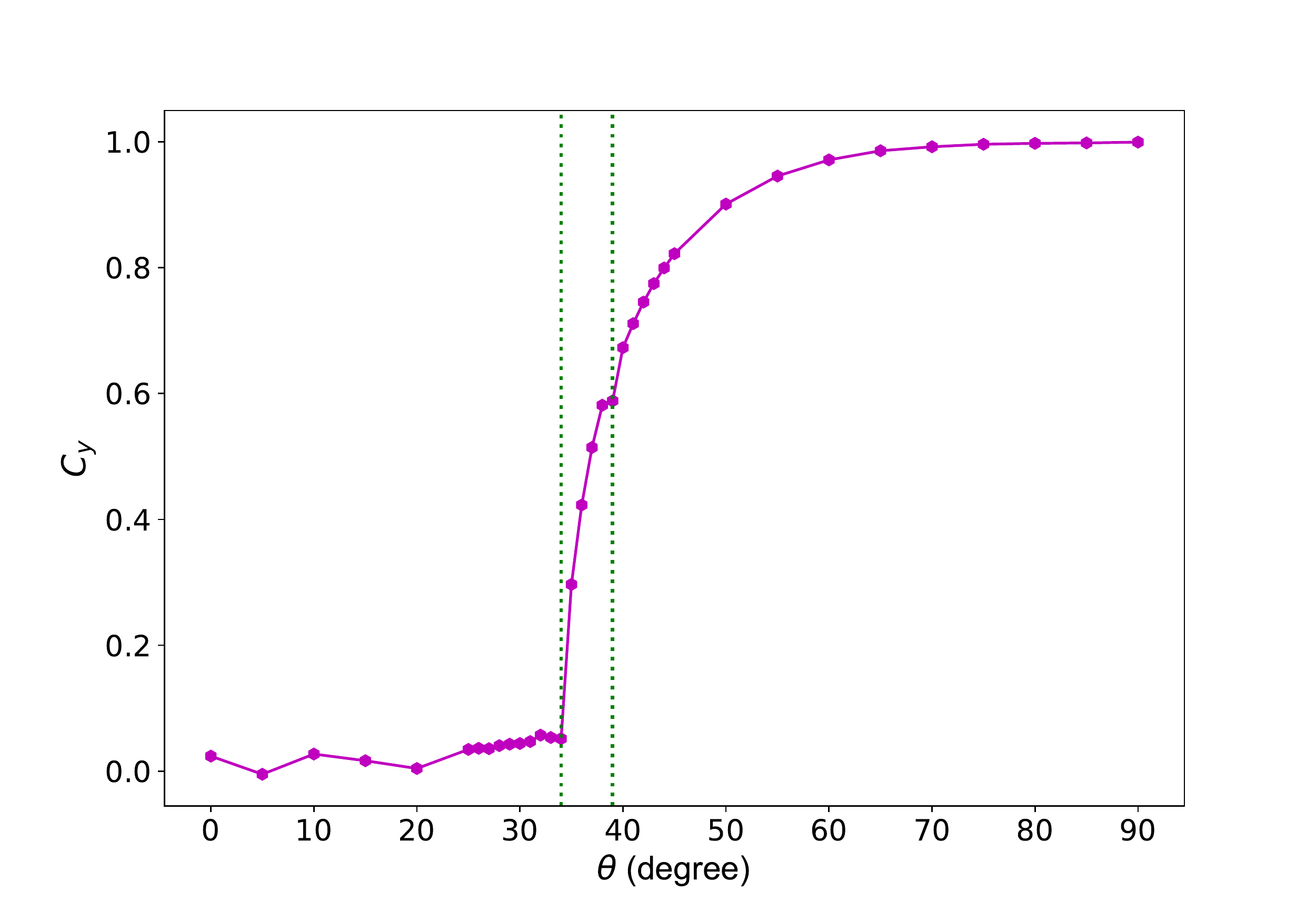}
\caption{The spin-spin correlation function $C_y$ defined in Eq. \eqref{Cy} as function of $\theta$.
It measures the out-of-plane antiferromagnetic order, and approaches $1$ in the Ising limit.
Its rapid decay to almost zero at $\theta=34^\circ$ {(indicated by the vertical dotted line on left)}
and the visible kink at $\theta=39^\circ$ {(vertical dotted line, right)}
suggest two quantum phase transitions.}\label{fig2}
\end{figure}

The first marker of phase transition is provided by  the spin-spin correlation function. 
It measures the antiferromagnetic long-range order and is defined by
{
\begin{equation}
C_y = \frac{4}{N-1}\sum_{\mathbf{r} \neq \mathbf{r}'}\eta_{\mathbf{r},\mathbf{r}'}\left[\langle S_y(\mathbf{r}) S_y(\mathbf{r}')\rangle - \langle S_y(\mathbf{r}) \rangle  \langle  S_y(\mathbf{r}')\rangle \right].
\label{Cy}
\end{equation}
Here $\mathbf{r},\mathbf{r}'$ label the sites, and $\eta_{\mathbf{r},\mathbf{r}'}=1$ (-1) if $\mathbf{r}$ and $\mathbf{r}'$ belong to the same (different) sublattice.} 
As shown in Fig. \ref{fig2}, $C_y$ decays from 1 in the Ising limit and 
drops sharply as the spin liquid phase is approached. The drop experiences a
glitch at $\theta=39^\circ$, and $C_y$ vanishes for $\theta\leq 34^\circ$.
This plot clearly demonstrates the existence of Neel order at $\theta>39^\circ$,
as well as the lack of out-of-plane antiferromagnetic correlation for $\theta< 34^\circ$ .

The second marker for phase transition is the expectation value of the in-plane spin 
{
\begin{equation}
S_\perp=\frac{1}{N}\sum_{\mathbf{r}} \sqrt{ \langle S_x(\mathbf{r}) \rangle ^2 + \langle S_z(\mathbf{r}) \rangle ^2}
\label{Sp}
\end{equation}
where $\langle S_x(\mathbf{r}) \rangle$ is the expectation value of the local spin operator $S_x(\mathbf{r})$,
and the sum $\sum_{\mathbf{r}}$  is over all the sites.}
Fig. 3 shows the variation of $S_\perp$ with $\theta$, which exhibits 
a trend opposite to $C_y$: it assumes large values within the bond ordered
phase, drops sharply around $\theta_1\sim 34^\circ$, followed by
a small glitch at $\theta_2\sim 39^\circ$. Afterwards, it remains suppressed and vanishes in the Ising limit
where all the spins align parallel to the $y$-axis. This result confirms that the spins are
predominantly in-plane within the bond ordered phase, in good agreement with TN results 
(see plot of $O_2$ in Fig. 2 of Ref. \cite{npj-tripod}). 
Combining Fig. 2 and Fig. 3 together, it is clear that an intermediate (the spin liquid) region is bounded by
the lower critical point $\theta_B=34^\circ$ and the upper critical point $\theta_N=39^\circ$. These critical values are close to, 
but not identical with, the TN results $\theta_B=33^\circ$ and $\theta_N=38^\circ$ 
(the phase boundaries were given in Ref. \cite{npj-tripod} in terms of $\theta'$, which
can be easily converted into $\theta$).

\begin{figure}[h]
\includegraphics[width=0.5\textwidth]{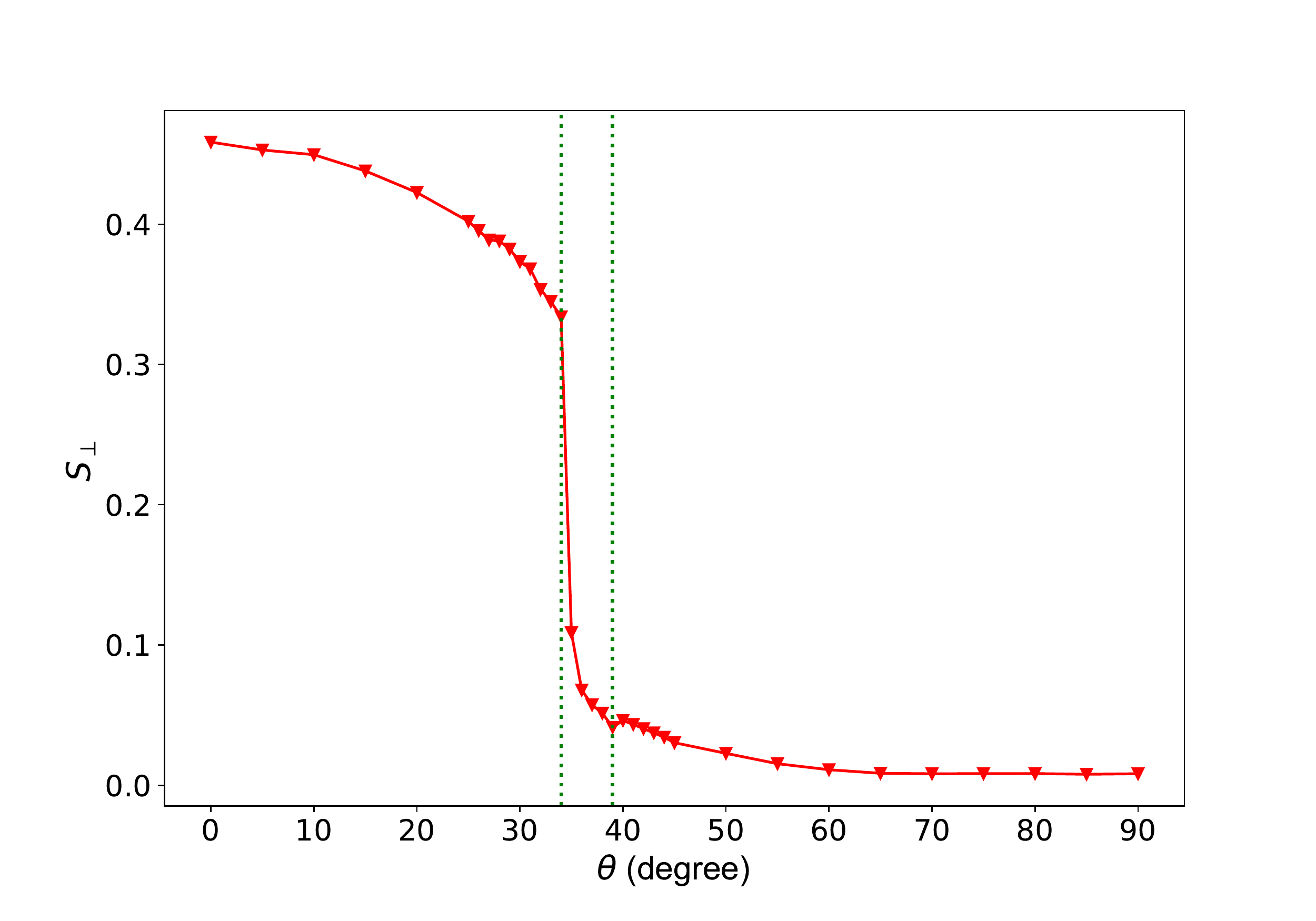}
\caption{The in-plane spin $S_\perp$ defined in Eq. \eqref{Sp}.
Its maximum is achieved at $\theta=0$ and remains finite within the bond ordered phase.
The sudden drop at $\theta=34^\circ$ provides a clear marker for the bond order to spin liquid transition.
Transition to Neel phase is accompanied by a small bump around $\theta\sim 40^\circ$.
Within the Neel phase, $S_\perp$ is very small and vanishes in the Ising limit.
}\label{fig3}
\end{figure}

To further elucidate the nature of the bond ordered phase, we compute the bond energies
\begin{equation}
B_\gamma (\mathbf{r})  =4\langle S_\gamma(\mathbf{r}) S_\gamma({\mathbf{r}+\mathbf{e}_\gamma})\rangle
\label{Bgamma}
\end{equation}
for all three bonds connected to a given site at $\mathbf{r}$. One then notices
that for $\theta<\theta_B$, one of the bond is stronger than the other two, and a bond modulation
pattern develops in space which breaks the underlying lattice symmetry.
There are three ways to break the symmetry of the three bonds locally. For example, in symmetry-breaking pattern $p_1$, 
$B_1$ is stronger (negative with larger magnitude)
while $B_2$ and $B_3$ are roughly (up to some small fluctuations) the same but weaker. 
The other two patterns $p_{2,3}$ are obtained by permuting $\gamma=1,2,3$, e.g. 
bond $B_2$ is stronger in pattern $p_2$. Obviously, these three patterns are related to each other by $C_3$ rotations in real space.
Let us define the bond modulation $\Delta B$ as the difference between the stronger bond and the average of the two weaker 
bonds, for example,
\begin{equation}
\Delta B= \frac{1}{N}\sum_\mathbf{r}\frac{1}{2}\left[ B_2 (\mathbf{r}) + B_3 (\mathbf{r})  \right] - B_1 (\mathbf{r}),
\label{deltaB}
\end{equation}
where the average over all sites is taken. A finite $\Delta B$ is expected if the bond modulation pattern $p_1$ is repeated 
throughout the lattice. As shown in Fig. 4, the bond ordered phase is characterized by a finite bond modulation, whereas
in both the Neel and spin liquid phase, bond energies are approximately uniform in space. 
Thus, the bond ordered phase found here 
has a solid order
of periodically modulated bonds, i.e. a valence bond solid. 
It breaks the $C_3$ symmetry of the underlying honeycomb lattice, but 
differs from the spin vortex state discussed in Ref.  \cite{npj-tripod}. 
And its energy per bond $\epsilon=-0.143$ is higher than the best TN result $-0.148$ \cite{npj-tripod}.
The reason behind this difference is addressed in 
the next section.

\begin{figure}[h]
\includegraphics[width=0.5\textwidth]{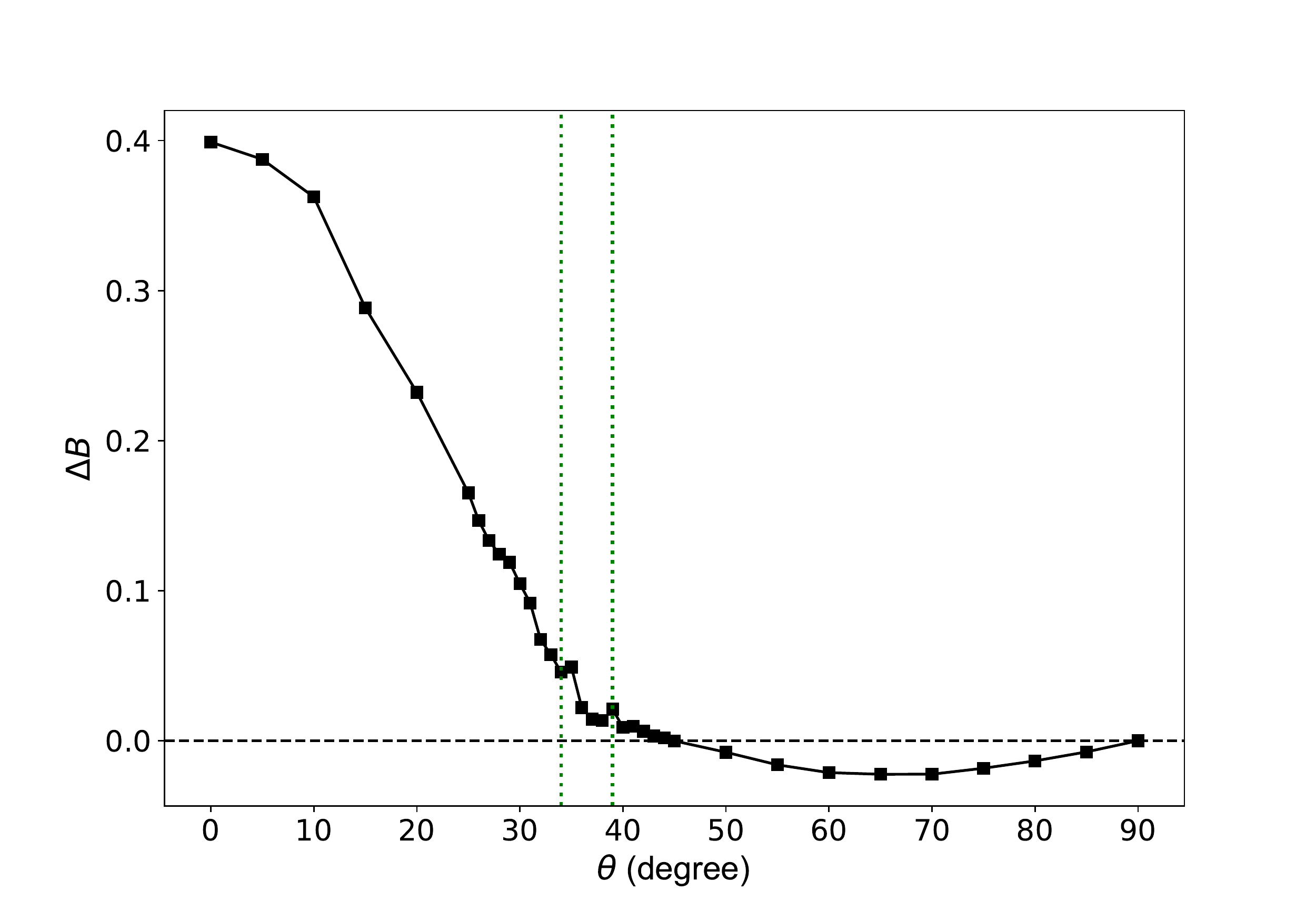}
\caption{Bond modulation $\Delta B$ defined by Eq. \eqref{deltaB} and \eqref{Bgamma}.
The bond ordered phase features bond modulations that grow with decreasing $\theta$. The Neel phase only
has small fluctuations in bond energy.
}\label{fig4}
\end{figure}

\section{Limitations and outlook}
To summarize, neural network quantum states based on RBM have performed very well to identify the main
phases (Neel, bond order, and spin liquid) and phase transitions of the frustrated tripod model. We find it remarkable that with 
some judicial control over the learning parameters and learning strategy, the algorithm can efficiently navigate the {$2^{32}$}-dimensional
Hilbert space stochastically to find variational ground states that have competitive energies.
In particular the two phase boundaries are 
close to the state-of-the-art TN ansatz. And for smaller systems, e.g. $L=3$, the energetics is also in excellent 
agreement with exact diagonalization.

Our study also exposes some limitations 
of the unconstrained RBM ansatz as implemented
here. 
In our calculation, we did not impose any symmetry constraints
on the RBM wave functions. While this has the advantage
of being completely unbiased, it also makes it exceedingly hard, if not at all impossible, to
reach intricate states such as the spin vortex lattice proposed in Ref. \cite{npj-tripod} for the 120$^\circ$ model. 
As shown in Fig. 4 above, our calculation reaches one of the valence bond states, where the 
bond modulation pattern $p_1$ is repeated periodically in space. 
There are two other states with degenerate energies, where pattern $p_{2}$ or $p_{3}$ is repeated instead. 
In fact, 
all possible coverings of the lattice by a suitable combination of local patterns $p_{1,2,3}$ have the same classical energy, giving rise
to a large residual entropy similar to those found in spin ice \cite{ice}. The quantum Hamiltonian Eq. (1) induces transitions
between different coverings and lifts the classical degeneracy. Then a particular covering, or superposition of coverings,
acquires lower energy to become the quantum mechanical ground state.
For example, the vortex state of Ref. \cite{npj-tripod} represents a
particular periodic covering of the whole lattice with local bond pattern $p_1$, $p_2$, and $p_3$.
In principle, this state may be eventually reached by RBM with further refinement in energy. In practice, this turns out to
be hard, due to the flatness of the energy landscape (since different covering patterns have close energies)
and the diminishing probability of settling into a highly symmetric configuration with a large unit cell by pure stochastic 
moves in a huge parameter space. 
Our attempts to further improve the energy 
frequently encounter numerical instabilities.
We conjecture that this barrier can be overcome by applying symmetry constraints \cite{choo-carleo-J1J2,nomura,symm}
to the RBM states, e.g. by enforcing $C_6$ symmetry and fixing the unit cell shape and size. 
This should also improve the convergence and numerical stability. The downside is that one must compare the energies 
of all candidate states with different symmetries. Testing this proposal is left for future work.

Fig. 4. illustrates another caveat of unconstrained learning in large systems: there are small but visible fluctuations in
the bond energy even in the Neel phase. For an ideal Neel state, one expects $\Delta B=0$.
While the algorithm successfully approaches the antiferromagnetic 
ground state with excellent energy, the RBM rarely settles into a completely frozen state with uniform bond energy.
The stochastic nature of the algorithm unavoidably introduces low lying excitations, which for larger systems 
are increasingly harder to eliminate. 
A similar situation is observed in Fig. 2, where the order parameter $C_y$ drops to almost zero within the bond ordered phase, 
but small fluctuations are still visible.
This presents a dilemma: on the one hand we need large clusters 
to accommodate orders with long modulation periods, on the other hand for large systems it becomes
more challenging to relax to pristine long-range ordered states.

Given these considerations, we advocate the following strategy to make the best out of the neural network ansatz.
First, the unconstrained network is trained to find the rough phase diagram and symmetry breaking patterns. It has the virtue of being
unbiased. Then, other methods, such as analytical variational wave functions or neural network with
symmetry, are used to further improve the energetics and elucidate the long-range order. We envision
such a hybrid approach will be especially useful in understanding complex spin systems, for example models 
inspired by a large class of Kitaev materials \cite{k-material-1,k-material-2}. Our results suggest that,
with further refinements
and complemented by other approaches, variational ansatz based on neural network quantum states 
can serve as a powerful tool to understand frustrated quantum spin models and more generally
strongly interacting many-body systems.


\begin{acknowledgments}
This work is supported by NSF Grant No. PHY- 2011386 (EZ). 
EZ would like to thank Christian McGuirk for preliminary work 
implementing and benchmarking RBM and Mahmould Lababidi,
Ahmet Keles and Haiyuan Zou for illuminating discussions. The numerical simulation is
based on the Netket library version 2.1.1 \cite{netket}. Part of the calculation was carried out on the ARGO 
clusters provided by the Office of Research Computing at George Mason University.

\end{acknowledgments}


\begin{thebibliography}{99}
\bibitem{fru1}C. Lacroix, P. Mendels, F. Mila, eds. Introduction to frustrated magnetism: materials, experiments, theory. Springer, 2011.
\bibitem{fru2} H. T. Diep, ed. Frustrated spin systems. World Scientific, 2013.
\bibitem{spin-liquid-review}L. Savary, L. Balents. Quantum spin liquids: a review. \emph{Reports on Progress in Physics} 80, 016502, 2016.
\bibitem{TN1} F. Verstraete and J. I. Cirac, Renormalization algorithms for quantum-many body systems in two and higher dimensions. \emph{arXiv:cond-mat}/0407066, 2004.
\bibitem{orus2014}R. Orus, A practical introduction to tensor networks: Matrix product states and projected entangled pair states. \emph{Annals of Physics} 349, 117, 2014.
\bibitem{tensor-spin}F. Verstraete, V. Murg, J. I. Cirac. Matrix product states, projected entangled pair states, and variational renormalization group methods for quantum spin systems. \emph{Advances in Physics} 57, 143, 2008.
\bibitem{mehta-rev}P. Mehta, M. Bukov, C.-H. Wang, A. G. Day, C. Richardson, C. K. Fisher, and D. J. Schwab. A high-bias, low-variance introduction to machine learning for physicists. \emph{Physics Reports} 810, 1, 2019.
\bibitem{bengio-repres}N. Le Roux and Y. Bengio. Representational power of restricted Boltzmann machines and deep belief networks. \emph{Neural Computation} 20, 1631, 2008.
\bibitem{carleo-troyer} G. Carleo and M. Troyer. Solving the quantum many-body problem with artificial neural networks. \emph{Science} 355, 602, 2017.
\bibitem{rbm-rev} R. G. Melko, G. Carleo, J. Carrasquilla, and J. I. Cirac. Restricted Boltzmann machines in quantum physics. \emph{Nature Physics} 15, 887, 2019.
\bibitem{choo-carleo-J1J2} K. Choo, T. Neupert, and G. Carleo. Two-dimensional frustrated J1-J2 model studied with neural network quantum states. \emph{Physical Review B} 100, 125124, 2019.
\bibitem{He-J1J2}X. Liang, W.-Y. Liu, P.-Z. Lin, G.-C. Guo, Y.-S. Zhang, L. He. 
Solving frustrated quantum many-particle models with convolutional neural networks.
\emph{Physical Review B} 98, 104426, 2018.
\bibitem{cai}Z. Cai and J. Liu. Approximating quantum many-body wave functions using artificial neural networks.
\emph{Physical Review B} 97, 035116, 2018.
\bibitem{sign2}M. Bukov, M. Schmitt, M. Dupont. Learning the ground state of a non-stoquastic quantum Hamiltonian in a rugged 
neural network landscape. \emph{SciPost Physics} 10, 147, 2021.
\bibitem{nomuraprx}Y. Nomura and M. Imada. 
Dirac-type nodal spin liquid revealed by refined quantum many-body solver using neural-network wave function, correlation ratio, and level spectroscopy. \emph{Physical Review X} 11, 031034, 2021.
\bibitem{brian}D. Kochkov, T. Pfaff, A. Sanchez-Gonzalez, P. Battaglia, and B. K. Clark. Learning ground states of quantum Hamiltonians with graph networks. \emph{arXiv}:2110.06390 (2021).
\bibitem{xiang-equiv}J. Chen, S. Cheng, H. Xie, L. Wang, and T. Xiang. Equivalence of restricted Boltzmann machines and tensor network states. \emph{Physical Review B} 97, 085104, 2018.
\bibitem{cirac-NN-TN} I. Glasser, N. Pancotti, M. August, I. D. Rodriguez, and J. I. Cirac. Neural-Network Quantum States, String-Bond States, and Chiral Topological States. \emph{Physical Review X} 8, 011006, 2018.
\bibitem{moore-TN}Y. Huang and J. E. Moore.  Neural Network Representation of Tensor Network and Chiral States.  \emph{Physical Review Letters} 127, 170601, 2021.
\bibitem{deng-entangle} D.-L. Deng, X. Li, and S. Das Sarma. Quantum Entanglement in Neural Network States. \emph{Physical Review X} 7, 021021, 2017.
\bibitem{deng-topo} D.-L. Deng, X. Li, and S. Das Sarma. Machine learning topological states. \emph{Physical Review B} 96, 195145, 2017.
\bibitem{chiral} R. Kaubruegger, L. Pastori, and J. C. Budich. Chiral topological phases from artificial neural networks. \emph{Physical Review B} 97, 195136, 2018.
\bibitem{imada} Y. Nomura, A. S. Darmawan, Y. Yamaji, and M. Imada. Restricted Boltzmann machine learning for solving strongly correlated quantum systems. \emph{Physical Review B} 96, 205152, 2017.
\bibitem{melko-fermion}P. Broecker, J. Carrasquilla, R. G. Melko, and S. Trebst. Machine learning quantum phases of matter beyond the fermion sign problem. \emph{Scientific Reports} 7, 8823, 2017.
\bibitem{khatami}K. Ch'ng, J. Carrasquilla, R. G. Melko, and E. Khatami. Machine learning phases of strongly correlated fermions. \emph{Physical Review X} 7, 031038, 2017.
\bibitem{sign1}C. Y Park, M. J. Kastoryano. Are neural quantum states good at solving non-stoquastic spin Hamiltonians? \emph{arXiv}:2012.08889, 2020.
\bibitem{npj-tripod} H. Zou, B. Liu, E. Zhao, W. V. Liu. A continuum of compass spin models on the honeycomb lattice. \emph{New Journal of Physics} 18, 053040, 2016.
\bibitem{kitaev} A. Kitaev. Anyons in an exactly solved model and beyond. \emph{Annals of Physics} 321, 2, 2006.
\bibitem{kita-rbm}M. Noormandipour, Y. Sun, B. Haghighat, Restricted Boltzmann machine representation for the ground state and excited states of Kitaev honeycomb model, \emph{Machine Learning: Science and Technology} 3, 015010, 2021.
\bibitem{rbm-MC}D. A. Puente, I. M. Eremin. Convolutional restricted Boltzmann machine aided Monte Carlo: An application to Ising and Kitaev models.
\emph{Physical Review B} 102, 195148, 2020.
\bibitem{xu-21}C.-X. Li, S. Yang, J.-B. Xu, Learning spin liquids on a honeycomb lattice with artificial neural networks,
\emph{Scientific Reports} 11, 16667 (2021) 
\bibitem{lode}N. Rao, K. Liu, M. Machaczek, L. Pollet. Machine-learned phase diagrams of generalized Kitaev honeycomb magnets. \emph{Physical Review Research} 3, 033223 , 2021 
\bibitem{compass}Z. Nussinov and J. van den Brink. Compass models: Theory and physical motivations. \emph{Review of Modern Physics} 87, 1, 2015.
\bibitem{ez-120}E. Zhao and W. V. Liu. Orbital order in Mott insulators of spinless p-band fermions. \emph{Physical Review Letters} 100, 160403, 2008.
\bibitem{wu-120} C. Wu. Orbital Ordering and Frustration of p-Band Mott Insulators. \emph{Physical Review Letters} 100, 200406, 2008.
\bibitem{nasu}J. Nasu, A. Nagano, M. Naka, S. Ishihara. Doubly degenerate orbital system in honeycomb lattice: Implication of orbital state in layered Iron oxide. \emph{Physical Review B} 78, 024416 (2008).
\bibitem{neutron}A. Scheie, K. Ross, P. P. Stavropoulos, E. Seibel, J. A. Rodriguez-Rivera, J. A. Tang, Yi Li, H.-Y. Kee, R. J. Cava, C. Broholm, 
Counterrotating magnetic order in the honeycomb layers of NaNi$_2$BiO$_{6-\delta}$. \emph{Physical Review B} 100, 214421, 2019
\bibitem{mit}I. Goodfellow, Y. Bengio, A. Courville. Deep learning. MIT press, 2016.
\bibitem{geron-book}A. Geron. Hands-On Machine Learning with Scikit-Learn, Keras, and TensorFlow: Concepts, Tools, and Techniques to Build Intelligent Systems. 2nd Edition, O'Reilly Media, 2019.
\bibitem{azizi}A. Azizi, M. Pleimling. A cautionary tale for machine learning generated configurations in presence of a conserved quantity. \emph{Scientific Reports} 11, 1, 2021.
\bibitem{netket}G. Carleo, K. Choo, D. Hofmann, J. E. T. Smith, T. Westerhout, F. Alet, E. J. Davis, S. Efthymiou, I. Glasser, S.-H. Lin, M. Mauri, G. Mazzola, C. B. Mendl, E. van Nieuwenburg, O. O'Reilly, H. Theveniaut, G. Torlai, A. Wietek. NetKet: A machine learning toolkit for many-body quantum systems. \emph{Software X} 10, 100311, 2019.
\bibitem{ice}A. P. Ramirez, A. Hayashi, R. J. Cava, R. Siddharthan, B. S. Shastry. Zero-point entropy in `spin ice'. \emph{Nature} 399, 333, 1999.
\bibitem{nomura}Y. Nomura. Helping restricted Boltzmann machines with quantum-state representation by restoring symmetry. \emph{Journal of Physics: Condensed Matter} 33,  174003, 2021.
\bibitem{symm}C. Roth and A. H. MacDonald. Group convolutional neural networks improve quantum state accuracy. \emph{arXiv}:2104.05085, 2021.
\bibitem{k-material-1}S. M. Winter, Y. Li, YH. O. Jeschke, R. Valenti. Challenges in design of Kitaev materials: Magnetic interactions from competing energy scales. \emph{Physical Review B} 93, 214431, 2016.
\bibitem{k-material-2}S. Trebst. Kitaev materials. \emph{arXiv}:1701.07056, 2017.
\end{thebibliography}
\end{document}